\documentclass[twocolumn]{aastex631}
\usepackage{amsmath}
\usepackage{graphicx}
\usepackage{placeins}
\usepackage{natbib}
\usepackage[caption=false]{subfig}
\usepackage{hyperref}
\usepackage{float}
\usepackage{indentfirst}
\defcitealias{ricarte2018observational}{RN18A}
\defcitealias{ricarte2018exploring}{RN18B}
\defcitealias{burke2025multiwavelength}{B25}
\newcommand{\RNA}{\citetalias{ricarte2018observational}}
\newcommand{\RNB}{\citetalias{ricarte2018exploring}}
\newcommand{\B}{\citetalias{burke2025multiwavelength}}

\begin{document}

\title{Constraining Supermassive Black Hole Seeding and Accretion via Occupation Fractions and Mass-Weighted Distributions}
\author[0000-0002-5680-1427]{Gianni Sims}
\affiliation{Florida Atlantic University, Department of Physics, Boca Raton, FL, USA}
\correspondingauthor{Gianni Sims}
\email{gsims2021@fau.edu}

\author[0000-0002-2304-0908]{Vicki Sarajedini}
\affiliation{Florida Atlantic University, Department of Physics, Boca Raton, FL, USA}

\shorttitle{}
\shortauthors{}
\shortauthors{}
\begin{abstract}
The formation of supermassive black holes (SMBHs) is a key open problem in galaxy evolution, with light- and heavy-seeding models predicting distinct black hole occupation fractions (BHOFs). Using multiwavelength observations from the LSST Deep Drilling Fields, we measure the AGN fraction and estimate the BHOF across $0<z<1$. Applying a hierarchical Bayesian model, we find that the inferred BHOF remains $100\%$ for host galaxy masses below $10^9 M_\odot$ at lower redshifts. These findings are consistent with light-seed scenarios. Additionally, we computed the black hole masses for several galaxies in the W-CDF-S to test the mass-weighted black hole occupation distributions from semi-analytic models (SAM). Using a reduced-$\chi^2$ analysis, we find that the data are consistent with the main-sequence accretion model regardless of the assumed seeding mechanism. These results provide new observational constraints on SMBH seeding and highlight the value of combining deep extragalactic surveys with probabilistic modeling. 
\end{abstract}


\section{Introduction}\label{1}
\begin{figure}[H]
    \centering
    \includegraphics[width = 8.5cm]{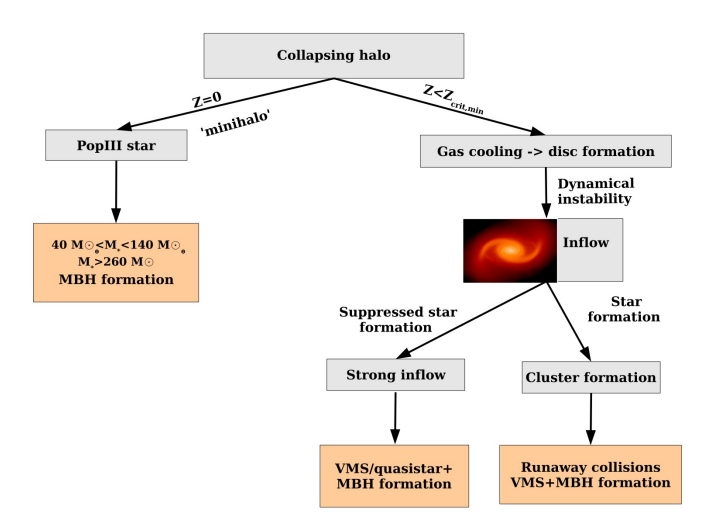}
    \caption{The flow diagram showcases the differences between the heavy and light seed models \citep{volonteri2010formation}.}
    \label{fig:1}
\end{figure}

While it is accepted that nearly all massive galaxies host a Supermassive Black Hole (SMBH) at their center, their presence in lower-mass systems remains an open question. This uncertainty is linked to another fundamental question: how did these SMBHs first form? There are two primary theories describing how SMBHs were formed (Figure \ref{fig:1}). In the light seed scenario, SMBHs formed from Population III stars. According to \cite{fryer2000pair}, Population III stars formed with high masses due to the inefficient cooling of molecular hydrogen gas in the early universe. Those with high initial masses leave behind black hole remnants of $\sim 100 M_\odot$. Some of the more massive stars, in the range of 140–260 $M_{\odot}$, explode as pair-instability supernovae, leaving no remnant behind \citep{heger2003massive}. \cite{madau2001massive} theorized that if these massive Population III stars formed in rare, high-density peaks at very high redshifts, they could produce a number density of BHs that matches the current number of SMBHs. However, if SMBHs were formed via the light seed model, they would require significantly higher or more sustained accretion rates. This often demands continuous Eddington or super-Eddington growth at high redshifts. Since the universe was less than a gigayear old during that time, these seeds had to grow exceptionally fast \citep{haiman2001highest, madau2014super, jiang2019super}.  Since light seeds originate from remnants of Population III stars, they start with lower initial masses and rely on the availability of gas and matter to accrete. In metal-free environments ($Z = 0$),
the formation of Population III stars occurs within minihalos.
These stars typically have masses ranging from $\sim100-1000 M_\odot$ and eventually collapse into light BH seeds of $\sim10-100 M_\odot$. In metal-poor environments ($Z < Z_{critical}$), the gas cools efficiently and leads to proto-galactic disc formation. This can either lead to suppressed star formation or star formation in clusters. Both processes contribute to SMBH formation. 

The second primary theory involves SMBHs being formed from the direct collapse of gas clouds, named the heavy seed model. This is only possible at very high redshifts due to the pristine gas needed to suppress cooling and fragmentation \citep{volonteri2010formation}. There are some stringent conditions that are required for this seeding mechanism to occur, such as the elevated Lyman-Werner background, additional sources of ionizing photons, and highly turbulent, rapid mass accretion with dense pro-galactic halos. Lyman-Werner (LW) background refers to the radiation in the 11.2 to 13.6 eV range, which can destroy molecular hydrogen (H$_2$) by dissociating it into atomic hydrogen. An elevated LW background will help heavy seeds to form because it prevents the gas from cooling efficiently by dissociating the molecular hydrogen, and the lack of fragmentation means that a large gas cloud can collapse directly into a massive black hole seed \citep{omukai2001primordial}. Another condition that could help heavy seeds to form is a process known as photoionization heating. In this situation, nearby star-forming regions or halos can emit a significant amount of ionizing radiation. This radiation can heat the surrounding gas, preventing it from cooling and fragmenting. Since the gas cloud can remain intact because of this process, it can then collapse into a single massive BH seed \citep{yoshida2003simulations, wise2019formation}. However, the exact number density of halos where this process could occur remains highly uncertain due to varying predictions on gas accretion rates and the required ionizing background \citep{inayoshi2020assembly}. The conditions required to form heavy seeds include high gas content but strictly low metallicity. Heavy seeds can only form in specific metal-free haloes at high redshifts since high-density, rapidly star-forming environments undergo fast metal enrichment from early supernovae. Consequently, many dwarf galaxies starting from $10^9 M_\odot$ remain isolated throughout their history and are left without heavy seeds. The heavy seed model allows for faster BH growth due to the larger initial mass of Very Massive Stars (VMS).

Theoretical models that follow the evolution of SMBHs have been developed to test the various seeding scenarios. \cite{volonteri2008evolution} created a Monte Carlo merger tree model which simulates the growth of black holes over time, providing a framework in predicting how many galaxies would host SMBHs. Based on those models, \cite{volonteri2010formation} constructed the Black Hole Occupation Fraction (BHOF), which examines the fraction of galaxies at given a mass that contain a SMBH.  

\begin{figure}
    \centering
    \includegraphics[width = 8.5cm]{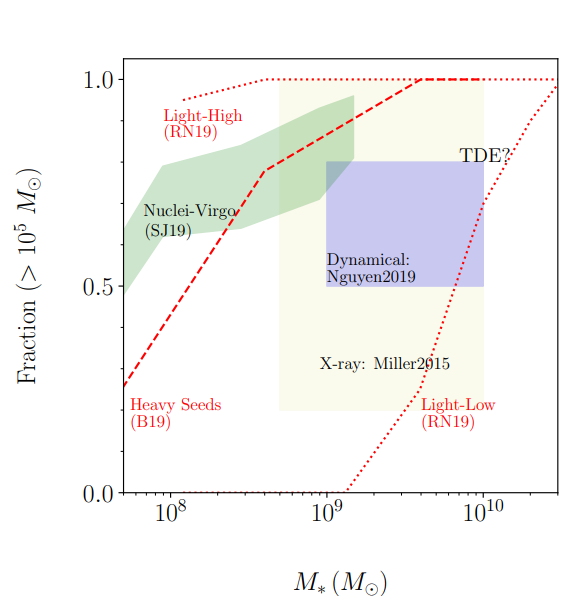}
    \caption{Theoretical BH seeding predictions for the BHOF compared to observational data \citep{greene2020intermediate}.} 
    \label{fig:2}
\end{figure}

\cite{ricarte2018observational} developed a semi-analytic model (SAM) embedded within the $\Lambda CDM$ framework to predict the BHOF for the different seeding mechanisms. The model simulates cosmic evolution by generating Monte Carlo merger trees for dark matter haloes across a wide mass range ($10^{11}$ to $10^{15}  M_{\odot}$). Within these trees, they implemented the two seeding mechanisms at high redshifts ($15 < z< 20$). As the haloes evolve and merge over dynamical friction timescales, the model tracks BH growth through a merger-triggered ``burst mode" regulated by energy-driven wind feedback, and a continuous ``steady mode", using either an AGN Main-Sequence (MS) or a Power-Law (PL) distribution. The AGN MS model relates the SMBH growth to the host galaxy's star formation, ensuring that BHs accrete gas at a steady-state even in non-merging galaxies. In contrast, the PL model follows a declining power-law function, which assumes that BHs spend most of their time at low, inefficient Eddington ratios \citep[hereafter \RNA,\RNB]{ricarte2018observational, ricarte2018exploring}. By tracking these dynamical mergers, tracking gravitational wave recoil ejections, and accounting for intrinsic observational scatter, the model produces the distinct BHOF tracks shown in Figure \ref{fig:2}. 

Intrinsically, the light seed model predicts a BHOF of nearly $100 \%$ across dwarf galaxies as Population III stars formed ubiquitously in early mini-halos. However, the models shown in Figure \ref{fig:2} include only BHs that have detectable masses of $M_{BH} >10^5 M_\odot$, requiring light seeds to have successfully accreted enough mass to cross this threshold. In the light seed framework, ``light-low" and ``light-high" correspond to different accretion histories. The light-low model uses a PL Eddington ratio distribution that features a low accretion rate, where BHs would spend most of their time accreting gas slowly and inefficiently. Because dwarf galaxies rarely undergo major mergers and have shallower potential wells to retain gas, the BHOF for the light-low model drops rapidly at lower galaxy masses in the local Universe. The light-high model uses an MS accretion mode, which enforces a consistent growth floor and allows BHs to grow continuously even in quiet environments. Compared to the light-low BHOF, the light-high model maintains a higher occupation fraction, but it also drops below $100 \%$ for dwarf galaxies with masses less than $10^9 M_\odot$. \cite{bellovary2019multimessenger} simulated the heavy seeding scenario where black holes form with relatively large initial masses. Because heavy seeds do not require extensive subsequent accretion to remain observable, this scenario leads to significantly higher BHOFs, specifically in dwarf galaxies that otherwise struggle to grow light seeds. 

Currently, observations have not yet shown which seeding mechanism is correct because the models require precise measurements in dwarf galaxies where data is limited. Existing observational boundaries are represented by the shaded regions in Figure \ref{fig:2}. The yellow-shaded region represents an observational constraint from the X-ray observations of low-mass red galaxies \citep{miller2015x}. The blue-shaded region represents a lower limit on the BHOF of greater than $50\%$, derived from dynamical BH mass measurements in a sample of nearby dwarf galaxies \citep{nguyen2019improved}. The green-shaded region traces the occupation fraction of nuclear star clusters derived from the Virgo cluster \citep{sanchez2019next}. These models suggest that the BHOF is relatively high for galaxies with stellar masses between $10^9 M_{\odot}$ and $10^{10} M_{\odot}$. However, they lack precision at lower masses, which is where the seeding models diverge the most. 

The goal of this research is to use the large extragalactic surveys of primarily SED-selected AGN from \cite{zou2022spectral} to observationally constrain theoretical models of BH seeding and growth. In Section \ref{2}, we describe the data used to construct the AGN fraction. In Section \ref{3}, we discuss the derivation and correction of the AGN OF, along with the use of Bayesian techniques to construct the BHOF. In Section \ref{4}, we evaluate the mass-weighted BH  distributions to understand their growth history. Finally, we present our discussions and conclusions in Sections \ref{5} and \ref{6}.

\section{AGN Selection} \label{2}
We used data from \cite{zou2022spectral} for the Deep-Drilling Fields (DDFs; e.g., \cite{brandt2018active}, \cite{scolnic2018optimizing}). These fields were selected as targets for the Vera C. Rubin Observatory Legacy Survey of Space and Time. Each field has a wide coverage of 4.9 $deg^2$ with around 0.8 million sources for the Wide Chandra Deep Field-South (W-CDF-S), 4.9 $deg^2$ with around 1.2 million sources for X-ray Multi-Mirror Mission Large Scale Structure (XMM-LSS), and 3.4 $deg^2$ with around 0.8 million sources for the European Large-Area Infrared Space Observatory Survey-S1 (ELAIS-S1). These fields have photometric observations ranging from X-rays via the XMM-Spitzer Extragalactic Representative Volume Survey (XMM-SERVS; \cite{chen2018xmm}; \cite{ni2021xmm}) to the far-infrared (FIR), with optical and near-infrared coverage from the VISTA Deep Extragalactic Observations (VIDEO; \cite{jarvis2013vista}) survey. They used CIGALE (Code Investigating GALaxy Emission) v2022.0 to perform Spectral Energy Distribution (SED) fitting to measure several properties of galaxies, including the presence of an AGN \citep{boquien2019cigale, yang2020x, yang2022fitting}. They compiled a sample of reliable AGN candidates by applying the Bayesian Information Criterion (BIC) method to identify sources that required an AGN template for the SED fit. They also selected AGN based on their X-ray emission properties and mid-IR (MIR) colors. 

\begin{figure}
    \centering
    \includegraphics[width = 8.5cm]{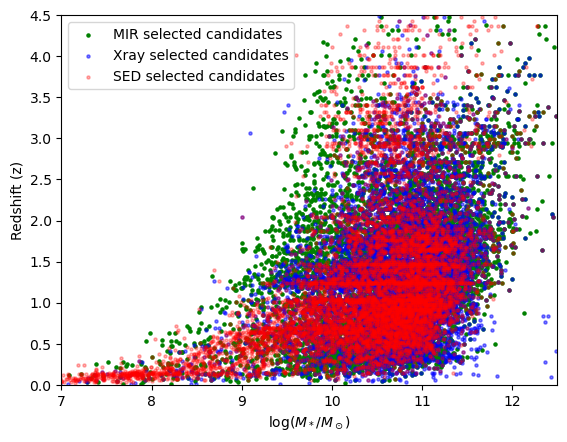}
    \caption{Redshift ($z$) as a function of stellar mass ($M_*$) for the AGN candidates in the DDF. Sources are color-coded by their corresponding selection methods: SED only (blue), X-ray only (red), and MIR only (green). This multiwavelength approach ensures a more complete selection of the AGN population across cosmic time.}
    \label{fig:3}
\end{figure}

Combining these high-purity SED, X-ray, and MIR criteria identifies a final sample of 8,364 AGNs in W-CDF-S, 7,999 in XMM-LSS, and 4,409 in ELAIS-S1. Among the SED-selected AGNs across all fields, 37$\%$ were also independently detected as X-ray sources, while 40.4$\%$ met the MIR color criteria, and 23.5$\%$ were identified by all three methods. Redshifts for this combined sample are derived from both spectroscopic and photometric measurements. Spectroscopic redshifts are available for $\sim29.5\%$ of the selected AGN population, while the remaining $\sim70.5\%$ rely on photometric redshift estimates. Figure \ref{fig:3} shows the redshift vs galaxy mass for the AGN candidates for all three fields. The data are color-coded by selection method, with overlapping multi-wavelength selections. 

\section{AGN and Black Hole Occupation Fractions} \label{3}
\begin{figure}[H]
    \centering
    \includegraphics[width = 8.5cm]{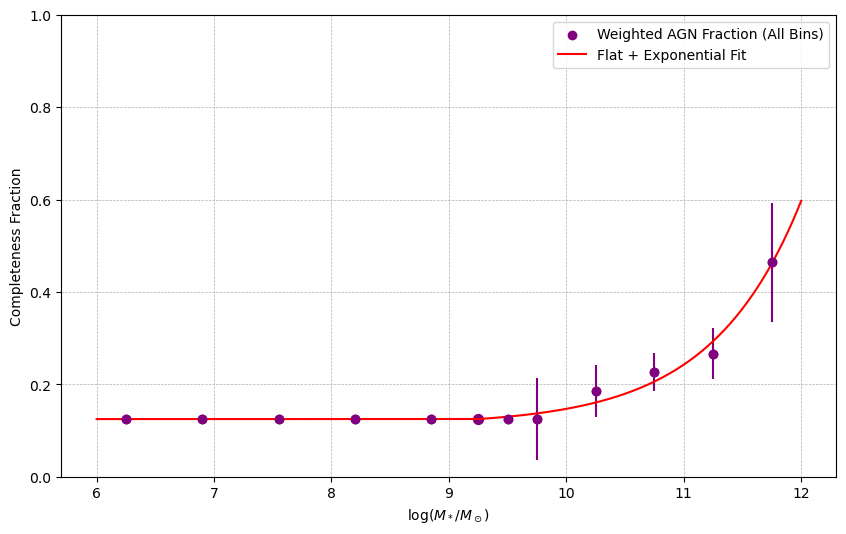}
    \caption{Completeness function based on the deep X-ray survey region with the W-CDF-S. Low-mass points were extrapolated from the lowest galaxy masses in the survey at $10^{9.25} M_\odot$.}
    \label{fig:4}
\end{figure}

Using this AGN sample, we first compute the AGN OF as a function of galaxy mass. To correct for incompleteness, we used the Chandra X-ray observations of the CDF-S \citep{luo2016chandra}. This 6-arcminute radius region within the W-CDF-S contains 345 X-ray sources. The 7Ms observations produce a largely complete and pure AGN sample. \cite{zou2022spectral} used this sample to tune their BIC criterion to identify the SED-selected AGN with the highest purity and completeness. Here, we use this sample to determine our incompleteness as a function of galaxy mass. Expanding on this method, we determined completeness by identifying which AGNs among the 345 deep-field X-ray sources remained detectable when applying the shallower X-ray flux limits of the entire W-CDF-S field ($\sim 1 \times 10^{-14} \text{ erg}\,\text{s}^{-1}\,\text{cm}^{-2}$). Above this flux limit, we would detect 36 of the SED-selected AGN, 74 of the MIR selected AGN, and 10 of the X-ray selected AGN. With some overlap between the detection techniques, this results in $\sim$ 20 percent of the truth sample being identified as AGN using these three detection methods. Separating this into galaxy mass bins produces the completeness function, where we limit ourselves to galaxies at $z < 1$ shown in Figure \ref{fig:4}. Since no AGN in the deep X-ray field were detected in galaxies with masses less than  $10^{9.25} M_\odot$, we extrapolate our lowest mass bin as a constant value of 0.125 for the range $10^6 M_\odot$ to $10^{9.25} M_\odot$.

\begin{figure}[H]
    \centering
    \includegraphics[width = 8.5cm]{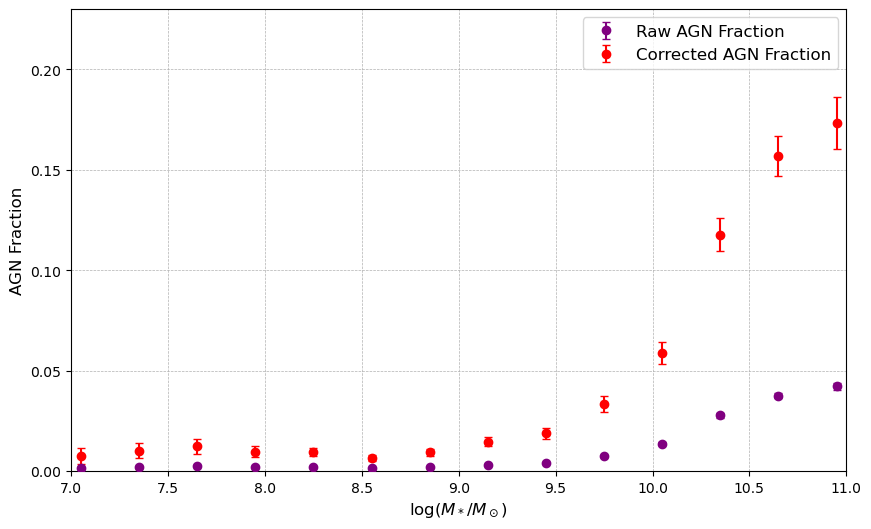}
    \caption{Using the correction function based on the deep X-ray observations, the observed AGN OF reaches 17$\%$ in the highest mass galaxies, dropping to only a few percent at masses less than $10^{9.5} M_\odot$.}
    \label{fig:5}
\end{figure}

Figure \ref{fig:5} shows our raw and corrected AGN OF. Applying the completeness corrections increases the raw fraction by a factor of $\sim 4.5$ for masses above $10^{9.5} M_{\odot}$ and by a factor of $\sim 2$ below this threshold. The noticeable drop in the AGN OF at lower stellar masses suggests that dwarf galaxies host a higher proportion of non-accreting BHs or do not host BHs at all. However, this low-mass population suffers from observational incompleteness because the faint accretion signatures may fall below the X-ray detection threshold.

\subsection{Estimating the BHOF} \label{3.1}
Several studies have attempted to infer the BHOF by taking into account detectability limits for the observations \citep{miller2015x, burke2025multiwavelength, zou2025central}. We adopt the technique of \cite{burke2025multiwavelength} (hereafter, \B), which utilizes a hierarchical Bayesian generalized linear model \citep{de2015overlooked} based on observed AGN luminosities that scale with the galaxy mass and observational upper limits. We focus only on the SED-selected AGN from the W-CDF-S for this analysis to maintain uniform photometric depth and selection completeness. 

\begin{figure*}[t]
    \centering
    \includegraphics[width = 19cm]{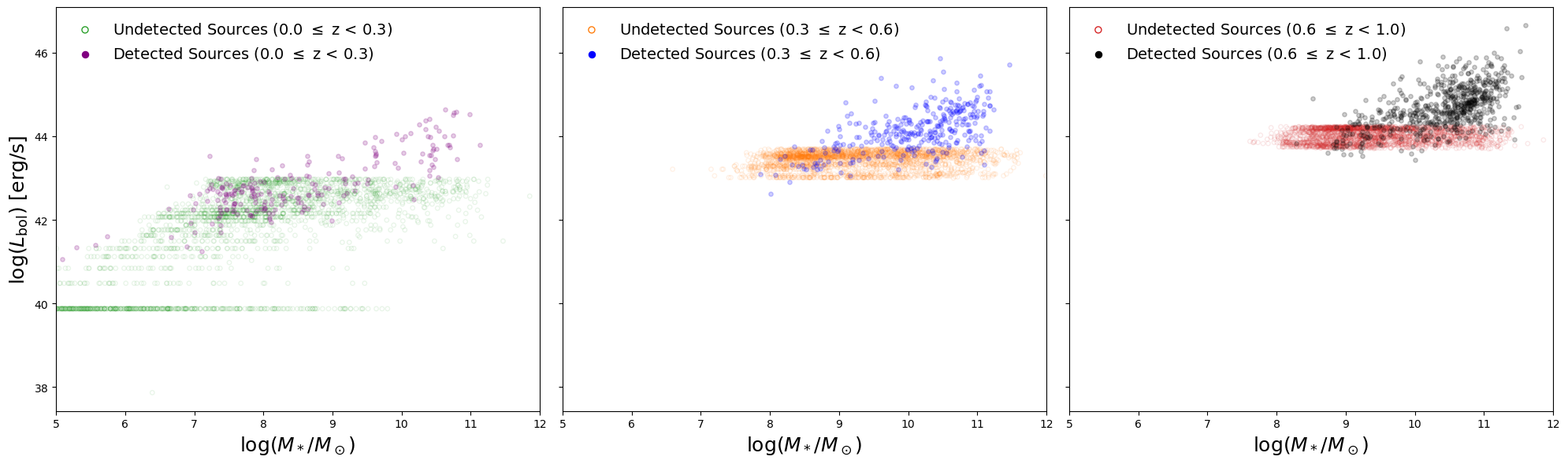}
    \caption{Each scatter plot shows the distribution of AGN bolometric luminosity as a function of host galaxy stellar mass in three redshift bins: $z \sim 0 - 0.3$ (left), $z \sim 0.3 - 0.6$ (middle), and $z \sim 0.6 - 1$ (right). Detected AGNs are shown as the purple, blue, and black points, respectively, while the smaller green, orange, and red points indicate the nondetection upper limits. As redshift increases, the upper limit luminosities increase due to larger luminosity distances.} 
    \label{fig:6}
\end{figure*}

We start by carefully determining the bolometric luminosities for both active and quiet BHs in our sources. Using the fitted SEDs, we derived the total bolometric luminosity by integrating the flux densities of the AGN component. To account for dormant sources that fall below the detection threshold, we generated an upper limits for the AGN nondetections. The minimum detectable luminosity is given by: 
\begin{equation}
    L_{lim}(z) = 4 \pi d_L(z)^2F_{lim} 
\end{equation}
where the flux limit of the survey, $F_{lim} = 3.141 \times10^{-14} \text{erg}\text{s}^{-1}\text{cm}^{-2}$, is defined as the fifth percentile of the observed flux distribution of the detected sample. This ignores the bottom $5\%$ of outliers and give a clean, uniform estimate of the flux limit. These non-detections are handled by marginalizing the probability distribution over all unobserved fluxes from $0$ up to $F_{\text{lim}}$. By explicitly integrating over these flux upper limits, the model naturally accounts for unobserved low-mass or low-luminosity systems without generating artificial drop-offs at the low-mass boundary. We divide the sample into three redshift bins: $z \sim 0 - 0.3$, $z \sim 0.3 - 0.6$, and $z \sim 0.6 - 1$. The $z \sim 0 - 0.3$ bin contains 216 detected sources and 102,063 nondetected sources, the $z \sim 0.3 - 0.6$ bin contains 321 detected sources and 102,734 nondetected sources, and the $z \sim 0.6 - 1$ bin contains 632 detected sources and 175,038 nondetected sources. Figure \ref{fig:6} shows the bolometric luminosity vs galaxy mass distributions across these redshift intervals.

\subsubsection{Computing BHOF Using Bayesian Modeling}\label{3.1.1}
We adopt the generalized logistic function used by \B\:
\begin{equation}
    BHOF(\log M_*) = \frac{1}{(1+\delta \exp(-\theta(\log M_* - \log M_{50})))^{\frac{1}{\delta}}}
\end{equation}
where $M_{50}$ represents stellar mass at which the occupation fraction reaches 50$\%$, $\theta$ controls the steepness of the transition, and $\delta$ provides additional flexibility in shaping the transition \citep{richards1959flexible}. 

To account for dormant BHs, we defined the detectable fraction $f_d$ as the probability that a galaxy has a BH that is both present ($f_{BHOF}$) and active ($f_a$). 
\begin{equation}
    f_d = f_{BHOF} \times f_a \times (1 - f_{obsc})
\end{equation}
For our analysis, we set $f_{obsc} = 0$. While a small population of Compton-thick sources may remain undetected, they are incorporated into $f_a$ and the intrinsic scatter of the model. This intrinsic scatter is defined by the $M_{BH}-M_*$ relation. Following \B\ and their adopted parameters from \cite{greene2020intermediate} for optically selected AGNs, we modeled the scaling coefficients as Normal distributions. We set the intercept to $\alpha_* \sim \mathcal{N}(6.70,0.13)$ and the slope to $\beta_* \sim \mathcal{N}(1.61,0.24)$. By doing so, we treated any heavily obscured, undetected BHs as statistically indistinguishable from an inactive one. This prevents overcomplicating the model while making sure that the variance caused by minor obscuration is still captured within the overall scatter of the model's accretion distribution.

To evaluate the model against the data, we used a likelihood function that treats detections and non-detections simultaneously. The total likelihood is given as:
\begin{equation}
\begin{split}
    \mathcal{L} = & \prod_{i \in \mathcal{A}_{\text{det}}} \left[ f_{\text{BHOF}}(M_{\star,i}) \, p(\log L_i \mid \log M_{\star,i}) \right] \times \\
    & \prod_{j \in \mathcal{A}_{\text{cens}}} \Big[ (1 - f_{\text{BHOF}}(M_{\star,j})) \\
    & \quad + f_{\text{BHOF}}(M_{\star,j}) \int_{-\infty}^{\log L_{\text{lim}}(z_j)} p(\log L \mid \log M_{\star,j}) \, d\log L \Big]
\label{eq:full_likelihood}
\end{split}
\end{equation}
where $\mathcal{A}_{\text{det}}$ represents the set of detected sources and $\mathcal{A}_{\text{cens}}$ represents the nondetected sources. The function in equation \ref{eq:formula} models the individual probability of observing an AGN with a given luminosity. The terms in the parentheses represent a Bernoulli probability distribution that accounts for the binary state of the BH being present or not. The Dirac delta term $\delta(log(L) + 9999)$ handles cases where the BH is either nonexistent, inactive, or heavily obscured. For a galaxy of a given mass, the probability of observing a specific bolometric luminosity depends on the specific Eddington Ratio Distribution Function (sERDF), which represents the probability distribution of accretion rates across the galaxy population. The intrinsic sERDF was modeled as a broken power law \citep{aird2012primus, aird2013primus, caplar2015agn, weigel2017agns, carraro2022eddington}, and then convolved with a Gaussian distribution to account for the intrinsic scatter in the BH-stellar mass relationship and observational uncertainties.  

\begin{multline}
    p(\log(L) | \log(M_*)) = p(I | \log(M_*)) [
    (f_d\space \mathrm{sERDF}(\lambda | \log(M_*))\\
    + (1-f_d) \delta(\log(L) + 9999))]
    \label{eq:formula}
\end{multline}

We used the prior distributions on the model parameters specified in \B\ (see their Table 1), and then explored the posterior distribution of the model parameters by using the No-U-Turn Sampler (NUTS) to efficiently sample the high-dimensional parameter space to obtain the best posterior estimates of the BHOF. 

\begin{figure}
    \centering
    \includegraphics[width = 8.5cm]{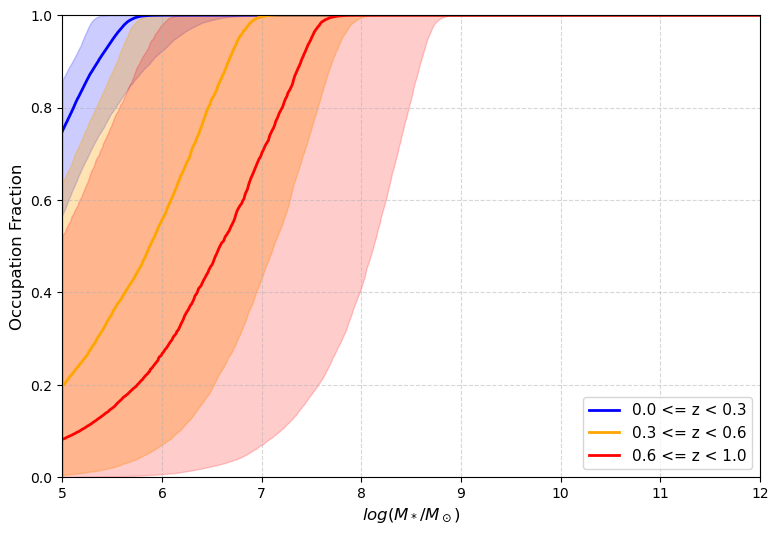}
    \caption{BHOF across three redshift bins: $z \sim 0 - 0.3$, $z \sim 0.3 - 0.6$, and $z \sim 0.6 - 1$ (left to right). Solid curves represent the median BHOF, while the shaded regions show the $1\sigma$ credible intervals.}
    \label{fig:7}
\end{figure}

The resulting BHOF curve for each redshift bin is shown in Figure \ref{fig:7}. Our results indicate that the occupation fraction is consistent with 100$\%$ for galaxies more massive than $10^7 M_\odot$ at all redshifts, with a greater decline in low masses at higher redshifts. We find that this decline at low masses corresponds to the mass limit for the galaxies in each of the redshift bins and therefore, the outcome differences for the BHOFs are unlikely to trace any evolution but are the result of incompleteness at the low mass end. 

To test this, we changed the $\log M_{50}$ prior for the intermediate and higher redshift bins to $U(7,12)$ and $U(8,12)$ to better reflect the range of galaxy masses sampled at each interval. Although the $1\sigma$ range for the intermediate and higher redshift bins narrowed, the BHOF was essentially unchanged and remained near $100\%$ to the lower galaxy mass limit of each redshift bin. This result is in agreement with \B, who similarly found a near-unity occupation fraction using a broad sample of X-ray, optical, and radio-selected AGN. The results align with the light seed model predictions. 

\section{Testing Mass-Weighted Occupation Distribution} \label{4} 
Theoretical predictions of BH seeding mechanisms also predict BH masses as a function of galaxy mass. \cite{ricarte2018observational} evaluated the mass-weighted occupation distribution for both seeding mechanisms with two different accretion methods to predict the BH mass distribution (see their Figure 4). They averaged the masses within given stellar-mass bins and divided them by the expected mass from the M-$\sigma$ relation. In this section, we test these predictions using BH masses measured for AGN in the W-CDF-S and other supplemental sources.

\subsection{Estimating the BH Mass} \label{4.1}
We cross-matched all 8364 SED, X-ray, and MIR-selected AGN from the W-CDF-S with spectroscopic catalogs from VLT/VIMOS and VLT/FORS2 \citep{balestra2010great, silverman2010extended, wuyts2008erratum}. We identified a total of 760 sources, consisting of 91 sources from FORS2 \citep{wuyts2008erratum}, 641 sources from VIMOS \citep{balestra2010great}, in addition to 28 X-ray-selected AGNs from \cite{silverman2010extended}. Out of this sample, we identified 29 that have broad-line emission features.  Distinguishing this broad-line subsample yields 13 sources with $\mathrm{C}\, \textsc{iv}$, 15 with $\mathrm{Mg}\,\textsc{ii}$, and 1 with $\mathrm{H}\, \beta$. For the remaining sources, the lack of an observed broad line does not mean an AGN is not present. In many cases, the broad emission lines fell outside the wavelength range of the observed spectrum. Some sources exhibit only narrow emission features, indicating the possibility of a narrow-line AGN or a host-galaxy dominated source. 

\begin{figure}[t]
    \centering
    \includegraphics[width = 9cm]{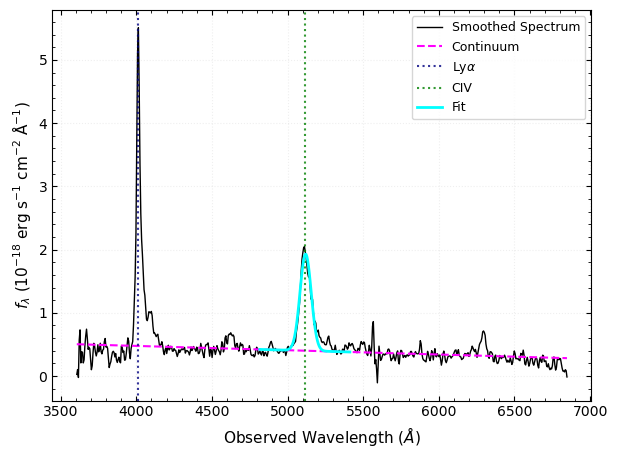}
    \caption{Example spectra decomposition of a broad $\mathrm{C}\,\textsc{iv}$ emission line for a source from the VIMOS sample. The linear continuum is represented as the magenta dashed line, and the cyan curve represents the best Gaussian profile for the broad emission feature. The expected rest-frame position of $\mathrm{C}\,\textsc{iv}$ is marked by the green dotted line, and the $\text{Ly}\alpha$ feature is marked by the navy dotted line.}
    \label{fig:8}
\end{figure}

Before spectral fitting, the spectra were smoothed using a boxcar moving average filter to reduce noise. Due to the difference in spectral resolutions between the archival datasets, both datasets were treated with different fitting strategies. Because VIMOS has a lower resolution ($R \sim 180\text{--}580$), emission lines and a linear continuum were fitted simultaneously using a single Gaussian profile to prevent overfitting and maintain numerical stability across the lower S/N multi-slit sample. To correct for this instrumental resolution, we quadratically subtracted the VIMOS instrumental broadening profiles from the observed width to recover the intrinsic velocity. For the higher-resolution FORS2 spectra ($R\approx660$), we performed a more detailed decomposition of the broad-line profiles. The local continuum was modeled from background sideband windows and subtracted before fitting. Emission features were modeled using a single Gaussian profile for narrow lines and a double-Gaussian profile for broad lines to accurately capture complex line wings. From these best-fit models, we extracted the resulting FWHM to measure the velocity of the broad-line region alongside the integrated line luminosity. Figure \ref{fig:8} shows one of the sources with a broad $\mathrm{C}\,\textsc{iv}$ emission line from the VIMOS subsample. 

We then estimated the central BH masses using the generalized single-epoch scaling relation:
\begin{equation}
\begin{split}
     \log \left( \frac{M_{\text{BH}}}{M_{\odot}} \right) = a + b \log \left( \frac{L}{10^{44}\ \text{erg\ s}^{-1}} \right) + \\
     2 \log \left( \frac{\text{FWHM}}{1000\ \text{km\ s}^{-1}} \right)
\label{eq:bhmass}
\end{split}
\end{equation}
where $L$ represents the monochromatic continuum luminosity ($\lambda L_\lambda$) matched to each respective line ($5100\text{ \AA}$ for $\mathrm{H}\, \beta$, $3000\text{ \AA}$ for $\mathrm{Mg}\,\textsc{ii}$, and $1350\text{ \AA}$ for $\mathrm{C}\, \textsc{iv}$). We adopt the calibrations of a = 6.91 and b = 0.50 for $\mathrm{H}\, \beta$, a = 6.66 and b = 0.53 for $\mathrm{C}\, \textsc{iv}$, and a = 6.86 and b = 0.50 for $\mathrm{Mg}\,\textsc{ii}$ \citep{vestergaard2006determining, vestergaard2009mass}. Our calculated BH masses range from $\sim$ $10^6M_\odot$ to $\sim$ $10^8M_\odot$. We also assumed a systematic uncertainty of $\sim 0.43$ dex due to intrinsic scatter in the virial factor calibration \citep{park2015cosmic, woo2010lick}.

\begin{figure}
    \centering
    \includegraphics[width=8.5cm]{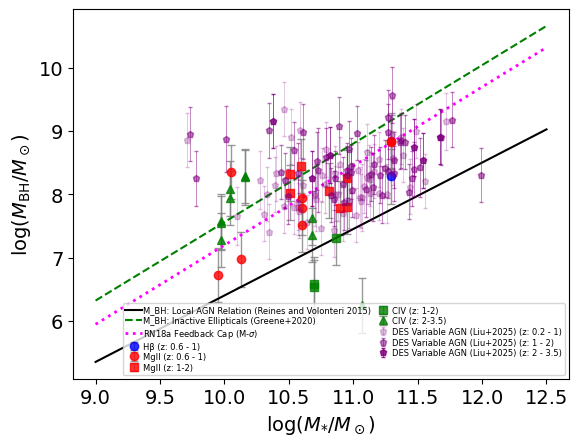}
    \caption{$M_{BH} - M_{star}$ relation for reliable AGNs with BH estimates. The data from this work are plotted as square, circle, and triangle symbols coded by redshift limits. In addition, emission lines are color-coded. The purple hexagonal points represent the sources from \cite{liu2025dwarf}, where redshift limits are coded by the transparency of the data point. The local $M_{BH} - M_{star}$ relation for inactive elliptical galaxies is plotted as a green dashed line \citep{reines2015relations}, the local relation for AGNs is plotted as a solid black line \citep{greene2020intermediate}, and the $M-\sigma$ relation used by \RNA\ is plotted as a magenta dotted line. We assume uncertainties of $\sim$ 0.43 dex on the BH masses \citep{park2015cosmic, woo2010lick}.}
    \label{fig:9}
\end{figure}

In Figure \ref{fig:9}, we show the BH vs galaxy mass for our 29 sources in W-CDF-S along with 111 variability-selected AGNs from DES \citep{liu2025dwarf},  which has a depth and redshift range comparable to the W-CDF-S. The sample exhibits significant scatter, falling between the local AGN relation and the relation for inactive ellipticals. Our sources are broadly consistent with the DES sample, with many of our higher-mass systems clustering near the inactive elliptical sequence. However, a subset of our sample measured by $\mathrm{C}\, \textsc{iv}$ and $\mathrm{Mg}\,\textsc{ii}$ scatters toward lower BH masses, aligning more closely with the local AGN relation. 

\begin{figure}[t]
    \centering   
    \includegraphics[width=8.5cm]{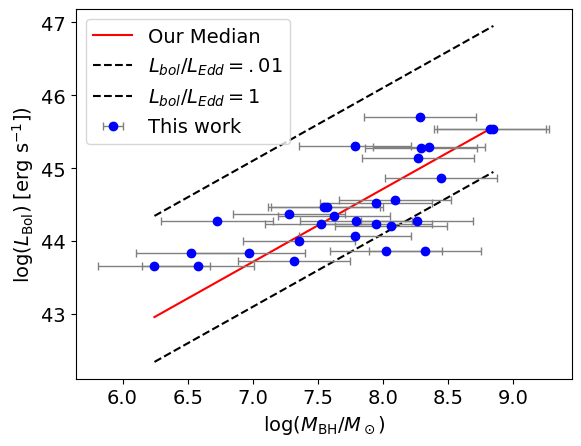}
    \caption{$L_{Bol} - M_{BH}$ Relation: The solid red line indicates the median Eddington ratio of our sample. For comparison, the black dashed lines show the Eddington ratios of 0.01 and 1. An uncertainty of $\sim 0.43$ dex is assumed for BH masses \citep{park2015cosmic, woo2010lick}.}
    \label{fig:10}
\end{figure}

To investigate the accretion properties of our sample, we plot the bolometric luminosity vs. BH mass for all 29 sources in Figure \ref{fig:10}, alongside the Eddington ratio corresponding to $\lambda_{\text{Edd}} = 0.01$ and $\lambda_{\text{Edd}} = 1.0$. The median Eddington ratio for our sources is $\sim0.04$, which aligns well with the median value of $\lambda_{\text{Edd}}\sim0.05$ for the DES AGN sample \citep{liu2025dwarf}. Other AGN samples, such as the X-ray selected AGN presented in \cite{suh2020no}, have a slightly higher median accretion rate of $\lambda_{\text{Edd}}\sim0.1$. This may indicate that SED-selected AGN are sensitive to lower accretion rate sources similar to those identified using variability selection.

\subsection{Mass-Weighted BH Distribution} \label{4.2}
We compare our sources and those of \cite{liu2025dwarf} to the predicted mass-weighted occupation distribution of \RNB\ (see their Figure 4). For each source, we computed the ratio of the measured BH mass to $M_\sigma$, where $M_\sigma = 10^{8.45}(\frac{\sigma}{200 \text{km s}^{-1}})^5 M_\odot$ (\RNA). In their models, \RNB\ constrained their AGN feedback parameters so that simulated BHs stop accreting upon reaching the $M-\sigma$ relation adopted by \cite{kormendy2013coevolution} at $z=0$. Because this relation excluded active emission-line AGNs, $M_\sigma$ represents the expected BH mass for a population of nonaccreting sources and serves as a theoretical normalization baseline. This is distinct from the empirical scaling relations for active AGNs and inactive BHs in Figure \ref{fig:9} and lies roughly between the two.  

\begin{figure*}[t]
    \centering
    \includegraphics[width = 19.5cm]{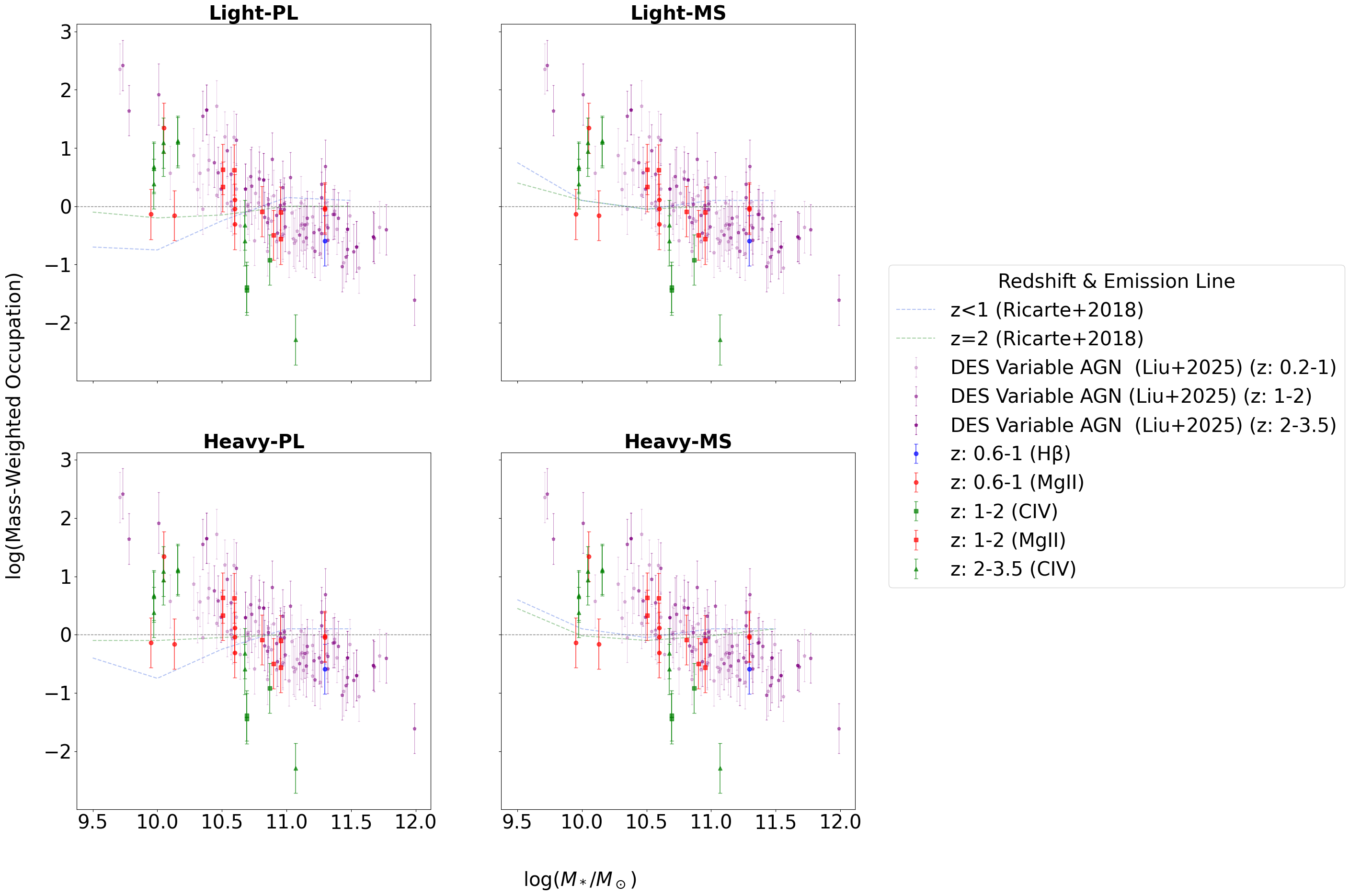}
    \caption{Comparison between our empirically constrained mass-weighted distribution and theoretical predictions from various SAMs. The colored points represent the VLT/VIMOS and FORS2 sources, and the dashed lines represent the approximate data points from \RNA. The different colors distinguish the redshift constraints, and the shapes of the points represent the emission lines used for mass estimates. An uncertainty of $\sim$ 0.43 dex is estimated for BH masses \citep{park2015cosmic, woo2010lick}.}
    \label{fig:11}
\end{figure*}

Figure \ref{fig:11} shows the mass-weighted BH distribution predictions of \RNA\ for PL and MS accretion modes for the light and heavy seed models as well as the data points from this study and \cite{liu2025dwarf}. The horizontal dashed line at $y=0$ represents BHs with masses predicted from the $M-\sigma$ relation. In PL models, SMBHs accrete using a universal PL Eddington ratio distribution to roughly reproduce the local luminosity function. In MS models, SMBH accretion rates and star formation rates (SFRs) are linearly correlated. 

 We performed a reduced-$\chi^2$ ($\chi_\nu^2$) analysis to determine which models best fit the data. The analysis reveals a preference for the MS accretion model, which yields a lower $\chi_\nu^2$ than the PL accretion model. The MS has $\chi_\nu^2 = 2.72$ for light seeds and $\chi_\nu^2 = 2.88$ for heavy seeds, while the PL has $\chi_\nu^2 = 3.65$ for light seeds and $\chi_\nu^2 = 3.49$ for heavy seeds. Although the data points show a large scatter around the theoretical models, this dispersion is within the 0.43 dex intrinsic scatter. As a result, the MS model remains a physically viable fit regardless of the different seeding mechanisms.

\section{Discussion}\label{5}
Our BHOF (Figure \ref{fig:7}) remains high ($f_{\text{occ}} \rightarrow 100\%$) extending to galaxy masses less than $10^9 M_\odot$, consistent with the local BHOFs presented in \B. The two results differ regarding the exact galaxy mass where the occupation fraction begins to decline. While \B\ observed the decrease to start around $10^7 - 10^8 M_\odot$, our derived $f_{occ}$ remains at $100\%$ down to much lower masses ($\sim 10^5 - 10^6 M_\odot$). However, we note that the vast majority of sources in our sample with masses below $10^7 M_\odot$ are upper limits with very few true detections to anchor the BHOF at low galaxy masses. \cite{miller2015x} used X-ray detected AGN from the AMUSE survey to establish a lower limit of $f_{occ}\sim 20\%$ for early-type galaxies below $10^{10} M_\odot$. Their results are also consistent with a high occupation fraction extended to dwarf galaxies, generally favoring light seed model predictions. However, recent results from \cite{zou2025central} are more consistent with the predictions from heavy seed models. They used a specific accretion ratio distribution function (sARDF) instead of sERDF to avoid assumptions about BH mass. The sERDF relies on empirical scaling relations, where accounting for intrinsic scatter in the host BH relations is essential for accurately modeling the occupation fraction. In contrast, the sARDF models the specific accretion rate as the ratio of X-ray luminosity directly to the host galaxy's stellar mass, which avoids the uncertainties of dwarf galaxy scaling relations. They also argued that their estimate is more reliable because they used deep Chandra X-ray observations of a large and highly complete sample of 1606 nearby galaxies. Their BHOF reveals a decline at host galaxy masses of less than $10^{10} M_\odot$. These conflicting results may be partly due to the different statistical approaches used to estimate the BHOF. While the sERDF assumes a broken power-law Eddington ratio distribution extending to low accretion rates, \cite{zou2025central} implemented a Box-Cox distribution for the sARDF that allows a turnover at low accretion efficiencies. Both approaches represent valid parameterizations of BH growth, illustrating how prior assumptions impact constraints on the low-mass occupation fraction.

In Figure \ref{fig:9}, we show that our sources are more closely aligned with the $M-\sigma$ relation for inactive ellipticals, consistent with the results for DES AGNs presented in \cite{liu2025dwarf}. Our data also reveal an abundance of overmassive BHs at the low-mass end of the galaxy distribution. While observational selection biases in flux-limited samples play a role by preferentially detecting more luminous systems near detection limits \citep{lauer2007selection, shen2010impact}, bias alone is unlikely to be the main cause of this low-mass overabundance. This overmassive population may be a consequence of early BH seeding combined with uncoordinated galaxy growth. \cite{mezcua2023overmassive} demonstrated that central BHs grow more efficiently than their host galaxies through early gas accretion and BH mergers at high redshifts ($z\gtrsim3$). While massive galaxies continue to grow through ongoing mergers and star formation to catch up to their central BHs, dwarf galaxies lack this late-stage growth and do not accumulate the stellar mass needed to align with the standard scaling relation. As a result, the overmassive dwarf population we observe is physically consistent with a population that preserves its early seeding history due to a lack of late-stage galaxy growth combined with observational biases.

We compare these sources and those of \cite{liu2025dwarf} to the mass-weighted occupation distributions of \RNA\ in Figure \ref{fig:11}. \RNA\ showed minimal differences among the seeding models in the mass-weighted occupation distributions. This could be due to cumulative accretion-driven growth dominating the present-day BH mass. This long-term growth obscures the initial seed mass, making the final distribution more sensitive to the accretion mode than seeding conditions. Although there is a large amount of scatter, the data are more consistent with the MS or ``high" accretion mode. A physical explanation could be that BH growth at $z < 1$ is predominantly driven by steady secular gas accretion that co-evolves with host galaxy star formation, rather than by merger-driven events. 
 
Combining our BHOF results favoring light seeds with the statistical preference for MS accretion is consistent with the ``light-high” seeding and evolution of black holes. In this scenario, light seeds can grow into the present-day SMBH population if their accretion is linked to the star formation history of their host galaxies.

\section{Conclusion}\label{6}
In this work, we investigated the formation scenarios of SMBHs by comparing observed AGN fractions and estimated BHOFs with predictions from theoretical seeding models. Using multiwavelength observations from the LSST Deep Drilling Fields \citep{zou2022spectral}, we find the following:
\begin{itemize}
    \item The AGN fraction for SED, X-ray, and MIR color selected AGN in the LSST DDFs reaches $\sim20\%$ for high mass galaxies but drops to just a few percent for galaxies with masses below $10^9 M_\odot$ and $z<1$.
    \item We used a hierarchical Bayesian framework based on the method presented in \B\ which assumes a power-law sERDF to model the detected AGN luminosities and the upper limits of nondetections. The resulting BHOF is consistent with $100\%$ occupation to low galaxy masses, consistent with the light seed models with a Main Sequence-like accretion mode, where the accretion is related to the star formation rate of the host galaxy. 
    \item We measured BH masses for a subset of the W-CDF-S sources and evaluated their mass-weighted distributions against theoretical SAM predictions for various seeding and accretion modes. The data are more consistent with the main sequence growth model, though the model predictions are redundant for different seeding mechanisms. Taken together with our BHOF results, the results of this study provide support for the light-high seeding and evolution models.
\end{itemize}

Future studies could revisit the DDF datasets using different assumptions about the accretion mode when calculating the BHOF. Observations that identify AGN in fainter, higher-redshift galaxies will also help distinguish among BH seeding models. Galaxy surveys with JWST and the future Roman observatory survey will allow us to understand the dwarf galaxy AGN population better and place better constraints on model predictions. 

\section{Acknowledgements}
We acknowledge the support from the LSST Data Science Fellowship Program, which is funded by many foundations such as the LSST Discovery Alliance, the Brinson Foundation, and the Gordon and Betty Moore Foundation. This work is based on multiwavelength observations across the LSST DDFs, along with data from the DES and spectroscopic observations from the European Southern Observatory (VLT/VIMOS and FORS2).

\clearpage
\bibliographystyle{aasjournal}
\bibliography{bib.bib}
\end{document}